\documentclass{agnspec}
\usepackage{graphics}
\usepackage{psfig}
\usepackage{./natbib}
\bibpunct{(}{)}{;}{a}{}{,}

\newcommand{\CIV}{\ion{C}{iv}}

\newcommand{\CIIIs}{\ion{C}{iii}]}

\newcommand{\cmsq}{\hbox{cm$^{-2}$}}

\newcommand{\nh}{\hbox{${N}_{\rm H}$}} 
\newcommand{\gtrsim}{\lower 2pt \hbox{$\, \buildrel {\scriptstyle >}\over {\scriptstyle\sim}\,$}}
\newcommand{\lesssim}{\lower 2pt \hbox{$\, \buildrel {\scriptstyle <}\over {\scriptstyle\sim}\,$}}
\newcommand{\asca}{{\emph{ASCA}}}

\newcommand{\chandra}{{\emph{Chandra}}}

\newcommand{\rosat}{{\emph{ROSAT}}}

\newcommand{\aox}{$\alpha_{\rm ox}$}

\newcommand{\balq}{BAL~QSO}
\newcommand{\balqs}{BAL~QSOs}
\newcommand{\etal}{et~al.}

\newcommand{\pgbal}{PG~2112+059}

\newcommand{\apj}{{ApJ}}
\newcommand{\apjs}{{ApJS}}

\newcommand{\aj}{{AJ}}
\newcommand{\aap}{{A\&A}}

\newcommand{\mnras}{{MNRAS}}
\begin{document}
\title{An Exploratory \chandra\ survey of Large Bright Quasar Survey broad absorption line QSOs}
\author{S.~C.~Gallagher,\inst{1} W.~N.~Brandt,\inst{1} G.~Chartas,\inst{1} \and\ R.~M.~Sambruna\inst{2}}
\institute{Department of Astronomy \& Astrophysics, The Pennsylvania State
University, 525 Davey Laboratory, University Park, PA U.S.A.
\and Department of Physics \& Astronomy and School of Computational Sciences,
George Mason University, 4400 University Drive M/S 3F3, Fairfax VA U.S.A.} 
\authorrunning{S.~C.~Gallagher~\etal}
\titlerunning{Exploratory \chandra\ survey of LBQS BAL~QSOs}
\maketitle
\begin{abstract}
We are in the process of obtaining a complete
sample of exploratory X-ray observations of broad absorption
line (BAL)~QSOs from the Large Bright Quasar Survey (LBQS).
This survey is composed of short (5--7~ks) \chandra\ ACIS-S3
observations designed to
significantly constrain X-ray fluxes and hardness ratios.
Our sample avoids the selection biases of some previous
surveys while providing X-ray data of similar quality.
Of 17 \balqs\ observed to date, 13 are detected. In general, our results are consistent
with absorption as the primary cause of X-ray weakness in \balqs.  
As found by \citet{GrEtal2001}, those \balqs\ with low-ionization absorption lines
are found to be notably X-ray weaker than \balqs\ with only high
ionization lines.
\end{abstract}
\section{Introduction}
\label{sec:lbqsintro}
From studies with \rosat\ and \asca, Broad Absorption Line (BAL) QSOs
are known to have faint soft X-ray fluxes compared to their
optical fluxes \citep[e.g.,][]{GrEtal1995,GrMa1996,GaEtal1999}
Given the extreme absorption evident in the UV, this soft X-ray faintness was
assumed to result from intrinsic absorption.  At the time, the
X-ray data could not explicitly demonstrate this, though 
the strong correlation found by Brandt, Laor, \& Wills (2000; hereafter
BLW)\nocite{BrLaWi2000} between \CIV\ absorption equivalent width (EW)
and faintness in soft X-rays supported this assumption.
The observation of \pgbal\ with \asca\ provided the first solid
spectral evidence
for intrinsic X-ray absorption and a normal underlying X-ray continuum 
in a BAL~QSO \citep{GaEtal2001a}.
Subsequently, more observations of BAL~QSOs with  \chandra\
also found signatures of intrinsic X-ray absorption \citep[e.g.,][]{GrEtal2001}.

\citet{GaBrChGa2002} compiled the results from the first moderate-sized 
survey of \balqs\ with enough counts for X-ray spectroscopic analysis.  They 
concluded from the spectroscopic evidence that the intrinsic
UV-to-X-ray spectral energy distributions of \balqs\ are
consistent with those of typical QSOs.  Furthermore, complex, intrinsic
absorption with $\nh=(0.1$--$4)\times10^{23}$~\cmsq\ is generally evident in the
X-ray spectra.  
At present, however, only a handful of these generally faint
targets have provided data of sufficient quality for spectral
analysis.  This sample is generally diverse, and thus far there
is no known predictor of observed 0.5--10.0~keV flux based on
the spectral properties in other wavelength regimes.  
In other words, there has been no obvious connection
between the characteristics of the UV and X-ray absorbers as
of yet, though all QSO with broad UV absorption also
apparently have X-ray absorption.

In an effort to remedy this situation, we are in the process of
compiling the multi-wavelength properties of a complete sample of
\balqs\ from the Large Bright Quasar Survey
\citep*[LBQS;][]{HeFoCh1995}. The sample will ultimately 
include all 37 of the \balqs\ from the LBQS with $z>1.35$, the redshift
at which the definitive \CIV\ BAL is shifted into the wavelength
regime accessible to ground-based spectroscopy.  
The optically bright targets are 
drawn from a homogeneous, magnitude-limited survey that has
effective, well-defined and objectively applied selection criteria
\citep*[e.g.,][]{HeFoCh2001}.  
One of the ultimate goals of this survey is to understand the connection
between UV and X-ray absorption in luminous QSOs, and hence gain some insight
into the mechanism for launching and maintaining energetic winds.  To
further this project, we are in the process of observing each object
in X-rays and in the rest-frame UV. 
In this paper, we present preliminary results from the analysis of the 
exploratory \chandra\
Advanced CCD Imaging Spectrometer (ACIS; 
G.~P. Garmire \etal, in preparation) observations of the first 17
targets which have been observed to date.
These short (5--7~ks) observations are intended to determine the basic X-ray
fluxes and rough spectral shapes of the sample, and for those targets
that were not detected, sensitive upper limits can be
set with these exposure times. In addition to
presenting new X-ray results, we also compare these data to some of the
multiwavelength information available in the literature.
All of the QSOs in this sample were observed in the
spectroscopic \balq\ survey of \citet[][hereafter WMFH]{WeMoFoHe1991}, and so 
a significant amount of data is available both from this work and
others based on the WMFH sample. 
\section{Observations and X-ray Data Analysis}
\label{sec:lbqsobs}
Each target was observed at the aimpoint of
the back-illuminated S3 CCD of  ACIS in faint mode.  The data were
processed using the standard \chandra\ X-ray Center
(CXC) aspect solution and grade filtering.  
Of the 17 observed \balqs, 13 were detected with counts 
ranging from 4--84 photons.  
After determining the position of the X-ray centroid, the counts in
the full (0.35--8.0~keV), soft (0.35--2.0~keV), and hard
(2.0--8.0~keV) bands were extracted from a source cell with a radius of
$2\farcs5$.  
The hardness ratio (HR), defined as the number of 2.0--8.0~keV counts divided by the
number of 0.35--2.0~keV counts, was then calculated; this parameter
provides a coarse quantitative measure of the spectral shape.
In order to transform the HR into a physical parameter, the photon index, for
characterizing the X-ray spectrum, observations of power-law X-ray
spectra were simulated using the  X-ray
spectral modeling tool, {\sc xspec} \citep{Arnaud1996}.  
For reference, a typical QSO with no intrinsic absorption and a photon index,
$\Gamma=2.0$, would have been observed to have HR=0.13--0.15 for the
range of Galactic column densities of (1.6--4.8)$\times10^{20}$~\cmsq\ 
in this sample.  
Once the value of $\Gamma$ corresponding to the observed HR and Galactic
\nh\ was determined, the model power-law spectrum was normalized using the
full-band count rate.  Given a photon index and normalization, the
derived X-ray properties, $F_{\rm X}$ (observed-frame 2--8~keV flux) and
$f_{\rm 2~keV}$ (flux density at rest-frame 2~keV) could be
calculated.  Finally, \aox, the slope of a hypothetical power law
connecting $f_{2500}$, the flux density at 2500~\AA, with $f_{\rm 2~keV}$, was determined.

\section{Testing for Significant Correlations}
Utilizing the available UV and optical data in the literature, we
sought evidence for correlations
that might indicate a connection between the X-ray properties and
those in other parts of the spectrum.  
To start, we considered the ``BALnicity index'' (BI), a conservative measure of the \CIV\
absorption equivalent width originally defined and tabulated by WMFH.
If the extent of X-ray weakness 
is due to intrinsic absorption, \aox\ might be expected to correlate with the
BI.  To test for such a relationship, we calculated the
non-parametric Kendall's $\tau$ statistic.
%
\begin{figure}[th]
\centerline{\psfig{figure=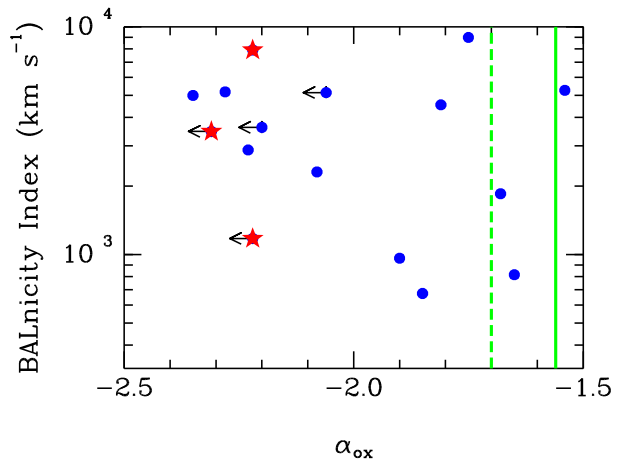,width=8.0cm}}
\caption[BALnicity Index vs. \aox]{BALnicity index vs. \aox.
The BI values are from
WMFH; this absorption-line parameter does not have a statistically significant
correlation with weakness in X-rays.
The solid and dashed vertical lines represent the mean and one standard deviation of
\aox\ for typical radio-quiet QSOs from \citet{BrEtal2001}. The stars
in this and subsequent plots indicate low-ionization \balqs.
\label{fig:bi}
\vspace*{-0.5cm}
}
\end{figure}
%

As can be seen in Figure~\ref{fig:bi}, there is no evident
correlation between the \CIV\ absorption-line BI and \aox;
this is supported by the high probability, $P=0.28$, that the null hypothesis
is consistent with these data.  This result could be considered
surprising given that \citet{BrLaWi2000} found a highly
statistically significant correlation between \CIV\ absorption
EW and \aox. However, their sample, the Bright Quasar Survey \citep{ScGr1983} 
objects with $z<0.5$, encompassed a much larger dynamic range of both \aox\ and
absorption EW; the majority of their sample exhibited neither weakness
in X-rays nor \CIV\ absorption.  Our \balq\ sample is probing the
most extreme end of \CIV\ absorption, where the straightforward
relationship between weakness in soft X-rays and UV continuum 
absorption apparently does not hold.  Given the well-documented complexity of
BAL profiles, which can include contributions from scattered flux as
well as exhibit severe saturation
\citep[e.g.,][]{OgCoMiTr1999,AravEtal2001}, the observed lack of
correlation is perhaps not remarkable.

\begin{figure}[th]
\centerline{\psfig{figure=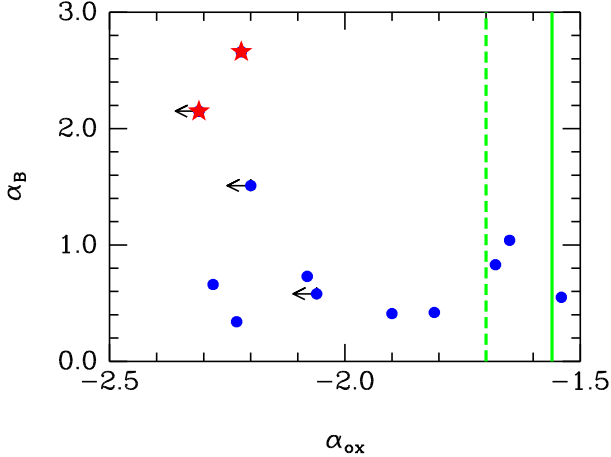,width=8.0cm}}
\caption[Blue continuum slope vs. \aox]{
The spectral index of the continuum blueward of \CIIIs, $\alpha_B$ for
$F_{\nu}\propto\nu^{-\alpha_B}$, versus
\aox.  Note that the reddest \balqs\ (those with the largest values of 
$\alpha_B$) are also X-ray weak.  The values for $\alpha_B$ are from
\citet{HuLaRe1998}.  
The vertical lines in the plot are as defined for
Figure~\ref{fig:bi}.
\label{fig:alpha}
\vspace*{-0.5cm}
}
\end{figure}

We also investigated possible correlations between the UV continuum
shape and X-ray properties.
Though the slope of the continuum blueward of \CIIIs, $\alpha_B$,
is not correlated significantly with \aox, there are some interesting
trends in the data (see Figure~\ref{fig:alpha}).  First, all of the
\balqs\ with the reddest continua (largest values of $\alpha_B$)
reside at the X-ray weak end of the sample.  However, not all of the
X-ray weakest \balqs\ are red.  If \aox\ can be associated with the
amount of intrinsic X-ray absorption, this suggests that the X-ray 
absorbing gas does not necessarily cause continuum reddening.
Physically this could result from absorbing gas with very low
gas-to-dust ratio as has been seen in Seyfert galaxies
\citep[][and references therein]{MaiolinoEtal2001}.  
Furthermore, dusty absorbers responsible for continuum
reddening may not be identical to those causing X-ray weakness.
%
%

\begin{figure}[th]
\centerline{\psfig{figure=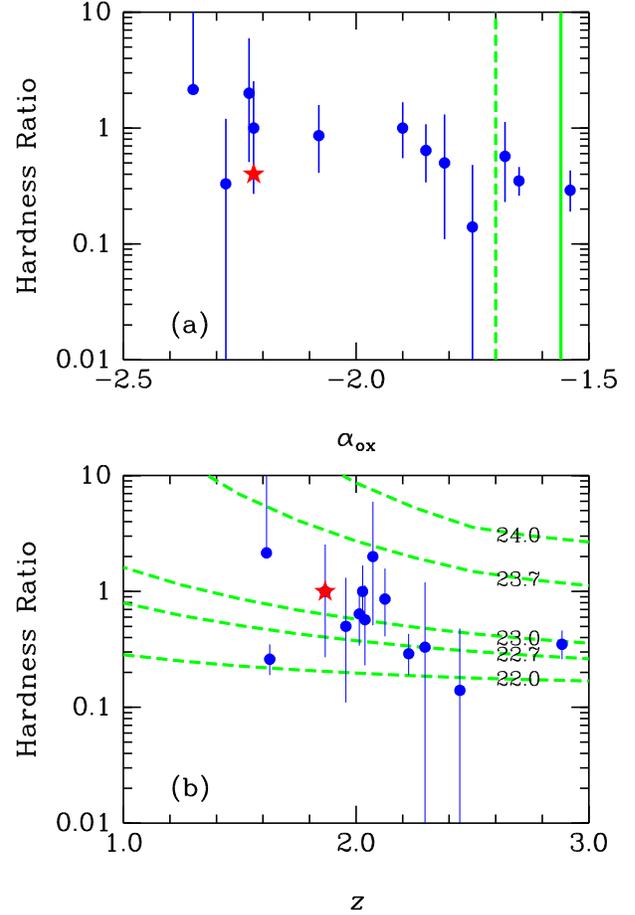,width=8.0cm}}
\caption[Hardness ratio plots]{
{\bf (a)} Hardness ratio versus \aox.  The weakest X-ray sources
(relative to their UV continuum flux) tend to have the
hardest spectra. 
The star symbols and vertical gray lines are as defined in the
caption to Figure~\ref{fig:bi}.
{\bf (b)} Hardness ratio versus $z$. 
The dashed curves represent the expected HR for a power-law continuum with
$\Gamma=2.0$, Galactic \nh$=2.8\times10^{20}$~\cmsq\ (the median value 
for this sample), and an intrinsic, neutral column density with the log value
labeled in the figure.  These curves are meant for reference only as
the actual absorption is likely to be complex \citep{GaBrChGa2002}.
\label{fig:hr}
\vspace*{-0.7cm}
}
\end{figure}

The search for correlations between absorption-line and
continuum properties with \aox\ did not result in any obvious
patterns, but those with the HR proved more fruitful.  
Though individual measurements of the HR suffer from large
statistical errors, the ensemble of HRs from the sample offers promise
for investigating the relationship of observed X-ray continuum shape
to other properties.  Tests of HR versus \aox\ 
resulted in a statistically significant correlation
(see Figure~\ref{fig:hr}a).
According to the evaluation of Kendall's
$\tau$, this correlation is significant at the $\gtrsim99\%$ 
confidence level. 
While typical QSOs have rest-frame 2--10~keV photon indices of $\Gamma=1.9\pm0.27$
\citep[e.g.,][]{ReTu2000}, a QSO exhibiting intrinsic absorption will 
have an apparently harder X-ray spectrum due to the lack of low-energy 
X-rays.  Thus, a larger value of the HR might result from intrinsic
absorption, which would also cause more negative values of \aox.

With these data, there is no statistically significant evidence for a correlation
between HR and $z$.  At higher $z$, the spectral
signatures of intrinsic absorption are pushed to lower observed
energies, and therefore occupy less of the soft X-ray bandpass.
An increase in HR with decreasing $z$ would therefore be
expected if all \balqs\ had similar amounts of intrinsic absorption.  
%
%
\section{Discussion}
The sample of \balqs\ presented in this paper offers the advantage of
being much more homogeneous than those in previous hard-band surveys
\citep[e.g.,][]{GaEtal1999,GrEtal2001}.
With the aim of obtaining the best X-ray constraints, these samples
included the optically brightest \balqs\ known, which were drawn
from surveys employing diverse selection criteria.  As a result, these 
objects were not necessarily representative of the global properties
of the \balq\ population as a whole.
In contrast, all of our targets are drawn from the same well-defined survey. 
In addition, our survey objects have uniform,
high-quality spectroscopic data from the rest-frame UV available for meaningful comparisons
between properties in different wavelength regimes.

From the days of \rosat\ and \asca, the detection fraction of \balqs\
has increased substantially. Of the 17 observed LBQS \balqs, 13 were
detected.  Given the sensitivity of \chandra, meaningful upper limits
on X-ray flux and \aox\ could be set for those that were not
detected.  For the observed values of \aox, our range encompasses
values from $-2.35$ to $-1.29$ with four upper limits and a median
value of $-2.06$.  For comparison, 
the mean and standard deviation for QSOs without evident UV absorption is
$-1.56\pm0.14$ \citep{BrEtal2001}.  
Not all of the \balqs\ are observed to have
\aox\ values that would group them as X-ray weak, e.g., with
\aox$<-1.7$.  While the calculated X-ray fluxes do suffer from large
errors due to the generally low count rates, \aox\ is not extremely
sensitive to these errors.  Another possible source of the spread
is variability of the UV continuum since some of these fluxes were
last measured more than 20 years ago \citep{HeFoCh1995}.
However, \aox\ is a robust parameter; for reference, an increase in
$f_{2500}$ of a factor of 2 would result in $\Delta$\aox$\approx0.1$.

From the X-ray data alone, we have evidence consistent with
the interpretation of intrinsic absorption as the cause of X-ray
weakness: the significant correlation of HR and
\aox.  An X-ray spectrum exhibiting intrinsic absorption by neutral,
partial-covering, or ionized gas will have fewer counts in the soft
band, and thus a higher value for the hardness ratio.  
At larger redshifts, the absorbed part of the spectrum will be shifted to lower
and lower energies thus occupying a smaller fraction of the observed
bandpass.  Therefore, for a given column density, an absorbed QSO at a 
lower redshift will have a higher value of HR than the same QSO at a
higher $z$.  This pattern is illustrated with the dashed lines in Figure~\ref{fig:hr}b.
However, making a straight identification of \aox\ with column density 
is problematic for two significant reasons.  First,
\citet{GaBrChGa2002} have shown that the intrinsic absorption is
complex, but with the present spectral resolution and signal-to-noise
ratio, the specific nature of that complexity is poorly constrained.  Thus the
assumption of neutral absorption, while useful for reference, is
probably not valid.  The effect of ionization state, covering
fraction, and velocity structure on measured column density from a CCD 
spectrum remains to be quantified.  Empirically, this complexity
manifests itself as additional flux at soft energies over what would
be expected from a completely neutral absorber. Given that situation, the HR and
\aox\ data presented in this sample should not be translated into
measurements of intrinsic column density.  

As found by \citet{GrEtal2001}, the low-ionization \balqs\ in this
sample (indicated by stars in Figures~\ref{fig:bi}--\ref{fig:hr})
also inhabit the X-ray weak end of \aox.  Attempting to correct 
the UV continua for dust extinction in these red QSOs would
only push \aox\ to more negative values. Given this evident
distinction in the X-ray as well as the UV properties, it is important
to distinguish the samples of low and high-ionization \balqs\ when trying 
to generalize about the population of \balqs\ as a whole.
\section{Future Work}
Many \balq\ studies have observed the most notable of objects.  While 
they frequently offer interesting case studies, this
situation can lead to spurious connections and inhibit
understanding of populations as a whole.
To avoid this pitfall, we are collecting exploratory X-ray observations 
of a complete sample of \balqs\ in conjunction with spectroscopic
X-ray data of the brightest ones.  By the end of 2002, we will have 5
additional exploratory observations.
Furthermore, we are in the process of
obtaining rest-frame UV photometry and spectroscopy for a
complete sample of LBQS \balqs, as described in
$\S$\ref{sec:lbqsintro}, to supplement the information in the literature as
well as the \chandra\ data.  
The imaging and spectroscopy, obtained with the
Hobby-Eberly Telescope within one month of each \chandra\ observation, 
will enable concurrent determinations of both $f_{2500}$ and $f_{\rm
2~keV}$ for measuring \aox, as well as the absorption-line
parameters.  Finally, these data will provide the means to investigate long-term
variability of the velocity structure of the BALs to test the
radiation-driven wind models of \citet*{PrStKa2000}.  In their
hydrodynamical models, they predict variability on timescales of a few 
years.  The most comprehensive \balq\ variability study to date, that
of \citet{Barlow1993}, was not of sufficient duration to test these models.  With 
the data of WMFH and our Hobby-Eberly Telescope spectroscopy, we can
approximately triple the timescale of \citet{Barlow1993} to span more
than a decade.

\begin{acknowledgements}
This work was made possible by \chandra\ X-ray Center grant GO1-2105X.
SCG also gratefully acknowledges support from NASA GSRP grant NGT5-50277 and
from the Pennsylvania Space Grant Consortium. 
\end{acknowledgements}

\end{document}